\documentclass[twocolumn,pra,superscriptaddress, amsmath,amssymb, aps,]{revtex4-1}

\usepackage{graphicx}
\usepackage{hyperref}
\usepackage{braket}
\usepackage{color}
\begin{document}

\title{Spin Hamiltonians are as hard as Boson Sampling}

\author{Borja Peropadre}
\affiliation{Department of Chemistry and Chemical Biology,
Harvard University, Cambridge, Massachusetts 02138, United States}
\author{Al\'an Aspuru-Guzik}
\affiliation{Department of Chemistry and Chemical Biology,
Harvard University, Cambridge, Massachusetts 02138, United States}
\author{Juan Jos\'e Garc{\'ia}-Ripoll}
\affiliation{Instituto de F{\'\i}sica Fundamental IFF-CSIC, Calle Serrano 113b, Madrid E-28006, Spain}

\begin{abstract}
Aaronson and Arkhipov showed that predicting or reproducing the measurement statistics of a general linear optics circuit with a single Fock-state input is a classically hard problem. Here we show that this problem, known as boson sampling, is as hard as simulating the short time evolution of a large but simple spin model with long-range XY interactions. The conditions for this equivalence are the same for efficient boson sampling, namely having small number of photons (excitations) as compared to the number of modes (spins). This mapping allows efficient implementations of boson sampling in small quantum computers and simulators and sheds light on the complexity of time-evolution with critical spin models.
\end{abstract}

\maketitle


Boson sampling requires (i) an optical circuit with $M$ modes, randomly sampled from the Haar measure; (ii) an input state with $N\ll M$ photons, with at most one photon per mode; (iii) photon counters at the output ports that post-select events with at most one photon per port. Under these conditions, the probability distribution for any configuration $\mathbf{n}\in\mathbb{Z}_2^{M}$ $p(n_1,n_2\ldots n_M)=|\gamma_{\mathbf{n}}|^2$, is proportional to the permanent of a complex matrix whose computation is \textbf{\#-P} hard. { This result, combined with some reasonable conjectures\ \cite{aaronson11}, implies that linear optics and interferometers have computing power that exceeds that of classical computation, and that the classical simulation of random optical circuits with non-classical inputs likely involves itself an exponential overhead of resources.} More recently, boson sampling has been generalized to consider other input states\ \cite{lund14,seshadreesan15}, extensions to Fourier sampling\ \cite{fefferman15} or trapped ion implementations\ \cite{shen14}. Boson sampling has also been related to practical problems, such as the prediction of molecular spectra \ \cite{huh14} and quantum metrology\ \cite{rohde15}. Finally, there are other quantum models, such as circuits of commuting quantum gates\ \cite{bremner15} which also establish potential limits of what can be classically simulated.

In this article we prove that boson sampling is equivalent to a many-body problem with spins that interact through a long-range, XY coupling { and evolve for a very short time, of the order of a single hopping or spin-swap event.} The model involves a bipartite set of input and output spins
\begin{equation}
H = \sum_{i,j=1}^M \sigma^+_{out,j} R_{ji} \sigma_{in,i}^- + \mathrm{H.c.},
\label{eq:spin}
\end{equation}
joined by the (unitary) matrix $R$. We show that the time evolution of an initial state that has only $N$ excited input spins approximates the wavefunction of the boson sampling problem after a finite time $t=\pi/2$. All errors in this mapping can be assimilated to bunching of excitations in the optical circuit, and the mapping succeeds whenever boson sampling actually does.

\begin{figure}[t]
\centering
\includegraphics[width=\linewidth]{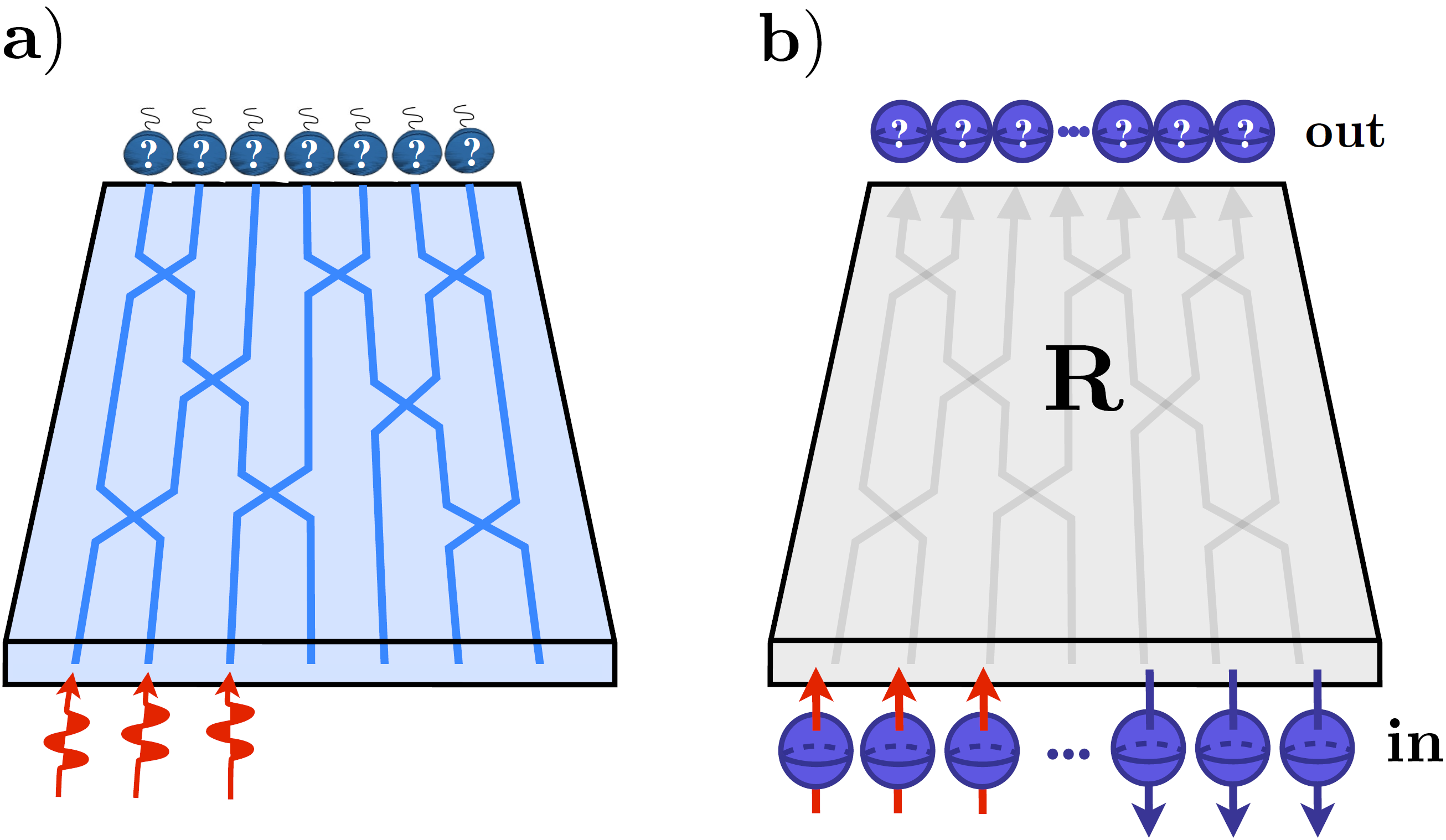}
\caption{(a) A setup that consists of beam splitters and free propagation implements boson sampling if the input state has a fixed number of single photons on each port (red wiggles). (b) We can regard those photons as arising from the spontaneous emission of two-level systems onto the circuit (up spins), which after propagation map onto other two-level systems at the end, via the unitary matrix $\operatorname {R}$.}
\label{fig:setup}
\end{figure}

Where does Hamiltonian (\ref{eq:spin}) come from? It arises from a setup with two-level systems as photo-emitters and detectors at the entrance and exit of an $M$-port interferometer [cf. Fig.\ \ref{fig:setup}b]. An interpretation of this setup in terms of earlier works with qubits in photonic waveguides\ \cite{chang06,gonzalez-tudela11,diaz-camacho15} shows it contains a coherent, photon-mediated interactions of the form (\ref{eq:spin}), accompanied by collective dissipation terms. In addition of this physical motivation \ \cite{olivares15},  our work introduces a new physical problem that has the same complexity as boson sampling, but which can be reproduced on state-of-the art quantum simulators and small quantum computers. Unlike boson sampling, spin sampling may be implemented using error correction, an idea that pushes the boundaries of what can be experimentally implemented beyond the limitations of linear optical circuits.

\section*{Results}
\subsection*{The boson sampling Hamiltonian}
Before studying the spin model\ (\ref{eq:spin}), let us develop a Hamiltonian for boson sampling. Let $R$ a the unitary transformation implemented by the circuit in Fig.\ \ref{fig:setup}a. $R$ is sampled from $\operatorname{U}(M)$ according to the Haar measure, and has associated an equivalent Hamiltonian
\begin{equation}
H_{BS} = \sum_{i,j=1}^M( b_j^\dagger R_{ji} a_i + \mathrm{H.c.})
+ \sum_{j=1}^M \omega(b^\dagger_jb_j + a^\dagger_ja_j).
\label{eq:harmonic}
\end{equation}
for the input and output modes, $a$ and $b$, of the problem. Evolution with this Hamiltonian transforms the initial state of boson sampling
\begin{equation}
\ket{\phi(0)} = a_1^\dagger \cdots a_N^\dagger\ket{\mathrm{vac}},
\label{eq:bs-input}
\end{equation}
into the $N$ boson sampling superposition
\begin{equation}
\ket{\phi(\pi/2)} = (-i)^N\prod_{i=1}^N \sum_j R^*_{ji} b_j^\dagger\ket{\mathrm{vac}},
\label{eq:bs-state}
\end{equation}
with a photon distribution given by the permanents $
|\gamma_{\mathbf{n}}|^2 = |\braket{\mathrm{vac}|b_1^{\dagger n_1}\cdots b_M^{\dagger n_M}|\phi(\pi/2)}|^2$, $n_i\in \{0,1\}$.

\subsection*{Dilute limit} 
\begin{figure}
\centering
\includegraphics[width=\linewidth]{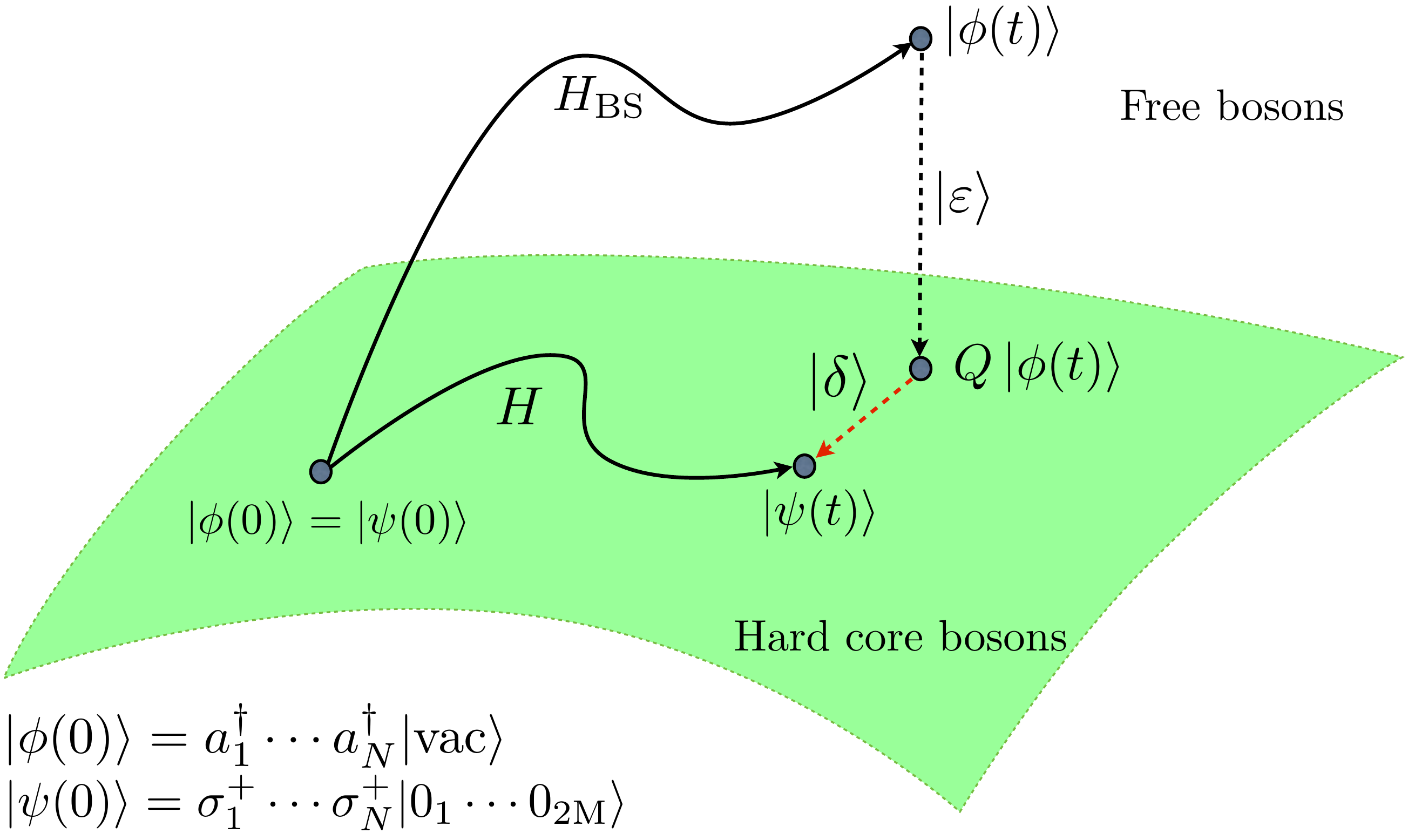}
\caption{The distance between the full bosonic state, $\ket{\phi(t)}$ and the state approximated with spins, $\ket{\psi(t)}$, is covered by two error vectors: one $\ket\delta$ that lives in the hard-core boson space (red dashed), and another one that covers the distance between the projected state $Q\ket{\phi(t)}$ and the full boson state, $\ket\phi$ (black dashed).}
\label{fig:errors}
\end{figure}

The final state in Eq.\ (\ref{eq:bs-state}) contains a non-zero probability of two or more bosons accumulating in the same mode. It can be split as in Fig.\ \ref{fig:errors}
\begin{equation}
\ket\phi = Q\ket\phi + \ket\varepsilon,
\end{equation}
where $Q\ket\phi$ is the projection onto states with zero or one boson per site and $\ket\varepsilon$ contains bunched states. For the errors $\ket\varepsilon$ to be eliminated in postselection while maintaining the efficiency of the sampling, the number of modes must be larger than the number of excitations. Formally the limit seems to be $M\simeq N^5\log^2(N)$, while $M\simeq N^2$ is the suspected ratio\ \cite{aaronson11} at which sampling becomes efficient, with bounds being tested theoretically and experimentally\ \cite{spagnolo13,arkhipov12}.

\subsection*{Spin sampling} 
The assumption of 'diluteness' of
excitations, which is needed for the efficient sampling of bosons, not
only ensures that we have a small probability of boson bunch at the
end of the beam-splitter dynamics, but also at all times. More
precisely, the probability of bunching of particles in the state
$\ket{\phi(t)}$, estimated by $\|Q{\phi(t)}\|_2^2$, roughly grows in time and
is bounded by the final postselection success probability [cf. Ref.\
\cite{epaps}]. In other words, boson sampling dynamics is efficient
only when it samples states with at most one boson per mode, the so
called hard-core-boson (HCB) subspace or the spin space. In this
situation one would expect that the models\ \ (\ref{eq:spin}) and
(\ref{eq:harmonic}) become equivalent, with the soft-boson corrections
becoming negligible.

Continuing with this line of thought, we now study how well the dynamics of the full bosonic system can be approximated by the hard-core boson model\ (\ref{eq:spin}). We regard the spin Hamiltonian as the projection of the full boson sampling model onto the hard-core-boson subspace, $H = Q H_{BS} Q$. Using this idea, we show that the boson-sampling dynamics is reproduced by this spin model at short times, with an error that grows with excitation density, and which is bounded by $\|\varepsilon\|^2$.

Let us assume that $\ket{\psi}$ is a hard-core boson state that initially coincides with the starting distribution of the boson sampling problem, $\ket{\psi(0)}=\ket{\phi(0)}$. This state evolves with the spin model as $i\partial_t \ket\psi = Q H_{BS} Q \ket\psi.$ Let us introduce the actual sampling error which we make by working with spins
\begin{equation}
\ket\delta = Q\ket\phi - \ket\psi.
\end{equation}
A formal solution for this error
\begin{equation}
\ket{\delta(t)} = -i\int_0^t e^{-i QH_{BS}Q (t-\tau)} Q H_{BS}
\ket{\varepsilon(\tau)}\mathrm{d}\tau.
\label{eq:delta-equation}
\end{equation}
Bounds for the different terms in this integral provide us with the core result (see supplementary material \ \cite{epaps}):
\begin{equation}
\|\delta(t)\|_2 \leq t \times \mathcal{O}\left(\frac{N^2}{\sqrt{M}}\right),
\label{equation_finalbound}
\end{equation}
which states that the error probability at $t=\pi/2$ is negligible when $N\sim { \mathcal{O}}	(M^{1/4})$. { In that material we also show numerical evidence\ \cite{epaps} that this bound can be at least improved to $N\sim { \mathcal{O}}(M^{1/3})$. Moreover, we show that the same bounds apply for the variation distance between the probability distributions associated to the quantum states $Q\ket{\phi}$ and $\ket{\psi}$, the measure used by Aaronson and Arkhipov in Ref.\ \cite{aaronson11}.}

Equation\ (\ref{eq:delta-equation}) shows that the errors due to using a spin model for boson sampling feed from bunching events in the original problem, $\ket\delta \propto \ket\varepsilon$, which are prevented by the HCB condition. For times short enough, these errors amount to excitations being ``back-scattered" to the ``in" spins. This means that sampling errors can be efficiently postselected in any given realization of these experiments, rejecting measurement outcomes where there contain less than $N$ excitations in the $\sigma^+_{out,j}$ spins. In this case, what we characterized as an error becomes a postselection success probability, $P_{ok}=1-\|\delta\|^2$.

\subsection*{Quantum computing}
Our previous results suggest an efficient implementation of boson sampling in a general purpose quantum computer using the following algorithm: (i) Prepare a quantum register with $M+M$ qubits, encoding the input and output spins, respectively. (ii) Initialize the register to the state $\ket{\psi(0)}=\ket{1_1,\ldots,1_N,0_{N+1},\ldots,0_{2M}}$. (iii) Implement the unitary $U=\exp(-i H\pi/2)$, where $H$ is given by (\ref{eq:spin}). (iv) Measure the quantum register. Postselect experiments where the $N$ qubits are at zero, recording the resulting state of the output qubits to estimate the sampling probability. Note that step (iii) may overcome the accuracy limitations of optical devices with the use of error correction, allowing scaling to arbitrarily large problems.
It is worth mentioning that while a general purpose quantum computer might implement boson sampling via the Schwinger representation of bosons, this would require larger number of qubit resources, and a greater complexity in implementing beam-splitting operations, while our spin sampling problem requires a smaller Hilbert space, and has a natural implementation on a small quantum computer.
\subsection*{Quantum simulation} We can use a quantum simulator with spins to implement spin sampling. As a concrete application, let us assume that we have a quantum simulator that implements the Ising model with arbitrary connectivity and coupling to a transverse magnetic field
\begin{equation}
H_{Ising} = \sum_{a,b=1}^{2M} J_{a,b} \sigma^x_a \sigma^x_b +
B \sum_{a=1}^{2M} \sigma^z_a.
\label{eq:ising}
\end{equation}

In the limit of very large transverse magnetic field, $|B|\gg\|J\|$, we can map this problem via a rotating wave approximation to the Hamiltonian\ (\ref{eq:spin}) where the coupling matrix is $J_{i,j+M}=J_{j+M,i}^*=R_{ij},\,(i,j=1,\ldots,M)$.

The Ising interaction\ (\ref{eq:ising}) is already present in trapped ions quantum simulators with phonon-mediated interactions\ \cite{porras04}, a setup which has been repeatedly demonstrated in experiments\ \cite{friedenauer08,kim09,islam11}, even for frustrated models\ \cite{kim10,islam13}, extremely large 2D crystals\ \cite{britton12}, and in particular in the XY limit\ \cite{jurcevic14}.
Another suitable platform for this kind of simulations is the D-Wave machine or equivalent superconducting processors with long-range tunable interactions \cite{Boixo14, Lanting14, Kelly15}. These devices can now randomly sample $J$ from a set of unitaries over a graph that is a subset of the available connectivity graph. Since the number of spins is very large, with over 900 good-quality qubits available, we expect that those simulations would surpass the complexity of the sampling problems that can be modeled in state-of-the art linear optics circuits.

\subsection*{Complexity theory} 
{ Our mapping of boson sampling to spin evolution shows that classically simulating the dynamics of long-range interacting spin models at short times has exactly the same complexity, which if the conjectures in Ref.\ \cite{aaronson11} hold, is \textbf{\#-P} hard. More precisely, if spin sampling were to be solvable in a classical computer, then we would approximate the boson sampling solution with precision poly($N^2/\sqrt{M}$), which would imply a collapse of the polynomial hierarchy \cite{aaronson11}. We have thus established a new family of problems that are efficiently simulatable in a quantum computer but not on a classical one.

This idea connects to earlier results that relate the difficulty of classically simulating time-evolution due to very fast entanglement growth\ \cite{badarson12,eisert06}. It also does not contradict the fact that free fermionic problems can be efficiently sampled because model\ (\ref{eq:spin}) only maps to free fermions for a subclass of matrices, $R$, which are tridiagonal.

There are other remarks to be done about our work and its place in the existing literature. First of all, it can be argued that spins or qubits are the underlying components of a quantum computer whose computation will in general amount to evolution with an effective Hamiltonian. This argument is bogus in that the resulting Hamiltonian will, in general, not be physically implementable, involving interactions to an arbitrary number of spins and distance. Moreover, even if certain models such \ref{eq:spin} are universal and may encode quantum computations\ \cite{kempe01,kempe02}, the timescales of our result amount to a single hopping event, which is scarcely the time to implement a single quantum gate and not an arguably complex computation. Finally, while at least one work has established connections between the collapse of the polynomial hierarchy and spin models\ \cite{bremner15}, that works builds on the conjecture that the complex partition function of a spin model is already in \textbf{\#-P}, and thus time-evolution of those spin models is hard to be approximated, which is instead the conclusion of this work.}

Summing up, we have established that boson sampling can also be efficiently implemented using spins or qubits interacting through a rather straightforward XY Hamiltonian. This map opens the door to simulating this problem with quantum simulators of spin model, of which we have offered two examples: trapped ions and superconducting circuits. Moreover, the same map states that boson sampling can be efficiently simulated in a general purpose quantum computer.

\section*{Methods}

\subsection*{Linear model} We build new orthogonal modes $c_j^\dagger = R_{ji} a_i^\dagger$, that satisfy the appropriate bosonic commutation relations, $[c_m,c_n^\dagger]=\sum_i R_{mi}^* R_{ni} = (R^\dagger R)_{m,n}=\delta_{n,m}$. These canonical modes almost diagonalize the previous Hamiltonian, which becomes a sum of beam-splitter models $H_{BS} = \sum_j (b^\dagger_j c_j + c_j^\dagger b_j).$
The dynamics of this Hamiltonian involves a swap of excitations from the normal modes $c_i$ into the output modes $b_i$, so that after a time $t=\pi/2$ the initial state\ (\ref{eq:bs-input}) is transformed into\ (\ref{eq:bs-state}).

It is very important to remark that the state $\ket{\phi(t)}$ can at all times be written as
\begin{equation}
\ket{\phi(t)} = \sum_{n=0}^N \cos(t)^n\sin(t)^{N-n}\ket{\xi_{n,M}},
\end{equation}
where $\ket{\xi_{n,M}}$ is the output state of a Boson-Sampling problem with $n$ input excitations in $M$ modes. Thus, $\ket{\phi(t)}$ has the bunching statistics of efficiently sampled Boson-Sampling problems, and it is ruled by the boson birthday paradox\ \cite{arkhipov12}. This in in sharp contrast with actual intermediate states in a random interferometer, which may contain a lot of bunching before the bosons exit the circuit, and it is due to the fact that the model that we use to recreate the BS output states, $H$, does not describe those intermediate stages, where the dynamics of individual beam-splitters matters.

\subsection{Spin model bounds}
Equation\ (\ref{eq:delta-equation}) arises from the simple Schr\"odinger equation
\begin{equation}
i\partial_t \ket\delta = Q H_{BS} Q \ket\delta + Q H_{BS}\ket\varepsilon.
\end{equation}
As explained above, it shows that the errors in approximating the boson sampling with spins result from the accumulation of processes that, through a single application of $H_{BS}$, undo a pair of bosons from $\varepsilon$, taking this vector into the hard-core boson sector.

We now bound the maximum error probability as an integral of two norms.  For that we realize that out of $\\varepsilon$, $Q H_{BS}$ cancels all terms that have more than one mode with double occupation. Thus,
\begin{equation}
\epsilon^{1/2}=\|\delta\|_2 \leq \int_0^t
\|Q H_{BS} P_{1bpair}\|_2 \|P_{1bpair}\ket{\epsilon(\tau)}\|_2
\mathrm{d}\tau,
\label{eq:spin-error}
\end{equation}
where $Q$ is a projector onto HCB states with $N$ particles and $P_{1bpair}$ is a projector onto the states with $N-2$ isolated bosons and 1 pair of $b$ bosons on the same site. As explained in the supplementary material\ \cite{epaps}, the value $\|P_{1bpair}\ket{\varepsilon(\tau)}\|^2=\|P_{1bpair}\ket{\phi(\tau)}\|^2$ is the probability of finding a single bunched pair in the full bosonic state. Combining a similar bound by Arkhipov\ \cite{arkhipov12} with the actual structure of the evolved state, we find
\begin{equation}
\|P_{1bpair}\ket{\epsilon(\tau)}\|_2 \leq \mathcal{O}\left(\frac{N}{\sqrt{M}}\right),
\end{equation}
which works provided that $N=\mathcal{O}(M^{3/4})$. We have also shown\ \cite{epaps} that the operator norm $\|Q H_{BS}P_{1bpair}\|_2$ is strictly smaller than the maximum kinetic energy of $N$ bosons in the original model, $H_{BS}$, so that
\begin{equation}
\|Q H_{BS} P_{1bpair}\|_2 \leq N.
\end{equation}
Combining both bounds we finally end up with\ (\ref{equation_finalbound}).

{ Note that because we only work with our bound for times $\Vert U\Vert_2 t\leq \pi/2$, the Boson-Sampling error $\ket{\varepsilon}$ contains only some multiply occupied sites at the out modes, $b_j$. Moroever, because of the construct in $\delta$, the dominant contribution from $\varepsilon$ to the error consists on emptying a single bunched site $b_j$ and using it to refill an input boson, $a_k$, leading to a spin configuration where an excitation has been ``back-scattered" to $\sigma_{in,k}$.}

%

\section*{acknowledgments}
B.P. and A.A.-G. acknowledge the Air Force Office of Scientific Research for support under the award: FA9550-12-1-0046. A.A.-G acknowledges the Army Research Office under Award: W911NF-15-1-0256 and the Defense Security Science Engineering Fellowship managed by the Office of Naval Research.
J.J.G.R. acknowledges support from Spanish Mineco Project FIS2012-33022, CAM Research Network QUITEMAD+ and EU FP7 FET-Open project PROMISCE.

\section*{Author contributions} B.P. and J.J.G.R. introduced the core idea and hypothesis; J.J.G.R. derived the final version of the proof; B.P., A.A.-G. and J.J.G.R. contributed to the scientific discussions and writing of the manuscript.

\clearpage
\begin{widetext}
\centering
\Large\textbf{Supplementary material}
\\
\end{widetext}

\section{Boson sampling dynamics}

Let us begin with the beam-splitter Hamiltonian
\begin{equation}
H_{BS} = \sum_j (b^\dagger_j c_j + c_j^\dagger b_j).
\label{eq:appendix-H}
\end{equation}
defined in terms of the transformed modes
\begin{equation}
c_j^\dagger = \sum_{i=1}^M R_{ji} a_i^\dagger.
\label{eq:relation}
\end{equation}
We need to study how the  evolution of an initial state with $N\ll M$ excitations
\begin{equation}
\ket{\psi} = \prod_{k=1}^N a_k^\dagger\ket{0},
\end{equation}
For that we write down the Heisenberg equations for  operators evolving as $O(t) = e^{-iHt} O e^{i Ht}$
\begin{align}
\frac{d}{dt} b^\dagger_j &= -i[H,b_j^\dagger]
= -i c_j^\dagger ,\\
\frac{d}{dt} c^\dagger_j &= -i b_j^\dagger,
\end{align}
which has as solutions
\begin{align}
b_j^\dagger(t) &= \cos(t) b_j^\dagger(0) -i\sin(t) c_j^\dagger(0),\\
c_j^\dagger(t) &= \cos(t) c_j^\dagger(0) -i\sin(t) b_j^\dagger(0).
\end{align}
Inverting the relation\ (\ref{eq:relation}), we recover
\begin{align}
a_k^\dagger(t) &= \sum R_{jk}^* c_j^\dagger(t)\\
&= \cos(t) R_{jk}^* R_{ji} a_i^\dagger(0)
-i \sin(t) R_{jk}^* b_j^\dagger(0)\nonumber\\
&=  \cos(t) a_k^\dagger(0)
-i \sin(t) R_{jk}^* b_j^\dagger(0),\nonumber
\end{align}
where we implicitly assume summation over repeated indices.
Dynamics under Hamiltonian\ (\ref{eq:appendix-H}) is coherently transferring population from the $a$ to the $b$ modes, as in
\begin{equation}
\ket{\phi(t)} = \prod_{k=1}^N \left(\cos(t)a_k^\dagger
-i\sin(t) R_{jk}^* b_j^\dagger\right)\ket{0}.
\label{eq:all-times}
\end{equation}
At time $t=\pi/2$ all population is transferred
\begin{align}
\ket{\phi(\pi/2)} &= \prod_{k=1}^N
a_k^\dagger(\pi/2) \ket{0},~\mbox{with}\\
a_k^\dagger(\pi/2) &= (-i)
\sum_{j=1}^M R_{jk}^* b_j^\dagger,\nonumber.
\end{align}
But in-between, $\ket{\phi(t)}$ may be regarded as a coherent superposition of different boson-sampling instances, $\xi_{BS,n}$ where only $n=0,1\ldots, N$ bosons participate and fed into the $M$ output modes, while $N,N-1\ldots,0$ remain in the input modes. In other words, we have
\begin{equation}
\ket{\phi(t)} = \sum_{n=0}^N  {N\choose{n}}^{1/2}\cos(t)^{N-n} \sin(t)^{n}
\ket{\xi_{BS,n}},
\label{eq:all-times-expansion}
\end{equation}
with normalized states $\ket{\xi_{BS,n}}$. 

It is important now to discuss each of these states, which we may write as
\begin{align}
\ket{\xi_{BS,n}}\propto \sum_{\{k,j\}}
a_{k_1}^\dagger(\pi/2)\cdots a_{k_n}^\dagger(\pi/2)\, \times &\\
\times \,a^\dagger_{j_1}(0)\cdots a^\dagger_{j_{N-n}}(0)\ket{0}&,\nonumber
\end{align}
and which consist on $N-n$ excitations that stay in the input modes $(a_j(0))$ and $n$ excitations that have been fully transferred to $a_k(\pi/2)$, which are linear combinations of the $b_k(0)$ modes. In other words, each of these instances $\ket{\xi_{BS,n}}$ represent themselves the outcome of a Boson Sampling experiments with $n$ input excitations and they will have the properties and statistics of bunching events of any boson sampling problem of such size\ \cite{arkhipov12}, a fact that we will use for proving bounds on the distance between $\ket{\phi(t)}$ and the hard-core-boson subspace at all times.
{
\subsection{Comparison with Optics}

It is very important to understand while the bunching statistics of $\ket{\phi(t)}$ does not contradict the behavior of actual optical circuits. In this work we are only studying the unitary transformation implemented by a boson sampling circuit. That transformation is generated by our toy Hamiltonian, $H_{BS}$
\begin{equation}
W_{BS} := \exp(-iH_{BS}\pi/2),
\end{equation}
but intermediate stages have no other physical reality than being an aid in proving our other results regarding the separation between boson and spin transformations. In particular, the optical transformation $W_{BS}$ will in general be implemented as a sequence of elementary transformations ---beam splitters and phase shifters---,
\begin{equation}
W_{BS} = \prod_i W_i,
\end{equation}
and intermediate stages of those transformations will, in absolute generality, lead to highly bunch and also highly entangled states which may be useful for other purposes. However, the final state $W_{BS}\ket{\phi(0)}$ can not be highly bunched, or otherwise it would not fit the framework of boson-sampling (And indeed it is not, following Arkhipov's boson birthday paradox).}

\section{Bunching bounds}
{
If we want Boson Sampling to be efficient, we need to impose that the number of bunching events in $\ket{\phi(\pi/2)}$ remains small with increasing problem size. Such property is guaranteed on average by the random unitaries $U$ sampled with the Haar measure, as explained by Arkhipov and Kuperberg in the boson-birthday paradox paper\ \cite{arkhipov12}. Below we will use the fact that the number of bunching events in $\ket{\phi(\pi/2)}=\ket{\xi_{BS,N}}$ indeed upper-bounds the number of bunches in each of its constituents, $\ket{\xi_{BS,n\leq N}}$, and use this idea to draw conclusions on the distance between the true Boson Sampling problem and the HCB spin model.

We cannot sufficiently stress the fact that the number of bunching events in $\ket{\phi(t)}$ is not related to the number of bunching events in the intermediate stages of a linear optics circuit. In order to implement a boson sampler one has to combine beam splitters that at different stages of the circuit cause the accumulation of the bosons. However, while those intermediate states in the construct are essential to reach the final Boson Sampling states, $\ket{\xi_{BS,n}}$, none of those intermediate states belongs to the family of states at the output of the circuit, $\ket{\xi_{BS,n}}$, which must have a low bunching probability (and which do, as shown by Ref.\ \cite{arkhipov12}).}

\subsection{One-bunching events}
For the purposes of bounding the error from the spin-sampling model, we need to bound the part of the error $\ket\varepsilon$ that contains a single pair of bosons on the same site, on top of a background of singly occupied and empty states. We have labeled that component $\|P_{1pair}\varepsilon\|_2^2$. {However, as discussed in Ref.\ \cite{arkhipov12}, bounding that probability is harder than bounding the probability $p_{HCB}(N,M)$ of having \textit{no bunching} event in a state with $N$ bosons in $M$ modes, distributed according to the random matrices $R_{ji}$}. This probability is
\begin{equation}
p_{\mathrm{HCB}}(N,M) = \prod_{a=0}^N \frac{M-a}{M+a} \simeq e^{-N^2/M}
\end{equation}
for dilute systems $N=\mathcal{O}(M^{3/4})$. We now use (i) that the state $\ket{\phi(t)}$ in Eq.\ (\ref{eq:all-times}) is made of a superposition of states with $n=0,1,\ldots N$ bosons distributed through the $M$ modes, (ii) that due to the randomness of $R$,{ each of these components shares the same statistical properties of the boson-sampling states}\ \cite{arkhipov12}, (iii) the probability distribution $p_{\mathrm{HCB}}(N,M)$ is monotonously decreasing with $N$. Using Eq.\ (\ref{eq:all-times-expansion}) and this idea we arrive at
\begin{align}
\|Q\phi(t)\|_2^2 &\simeq
\sum_{n=0}^N {N\choose{n}}\cos(t)^{2(N-n)}\sin(t)^{2n} p_{\mathrm{HCB}}(n,M)\nonumber\\
&\geq \sum_{n=0}^N {N\choose{n}}\cos(t)^{2(N-n)}\sin(t)^{2n} p_{\mathrm{HCB}}(N,M)\nonumber\\
&= (\cos(t)^2+\sin(t)^2)^N p_{\mathrm{HCB}}(N,M)\nonumber\\
&= p_{\mathrm{HCB}}(N,M).
\end{align}
Using the fact that $Q\ket\phi$ and $\ket\varepsilon$ are orthogonal and thus $\|\phi\|^2_2 = \|Q\phi\|^2_2+\|\varepsilon\|^2_2,$ we can find a very loose bound for the error probability of single bunching events
\begin{equation}
\|P_{1bpair}\varepsilon\|_2^2\leq
\|\varepsilon(t)\|_2^2 \leq 1 - p_{\mathrm{HCB}}(N,M).
\end{equation}
Note that this bound can be translated into an upper bound of $\mathcal{O}(N^2/M)$ using the fact that the exponential falls faster than $1-N^2/M$.

\section{HCB operator bound}

In addition to bounding the error vector, we also need to bound the norm of an operator that brings back population from the error subspace into the hard-core-boson subspace. Because $\|P_{1bpair}\varepsilon\|_2$ is already rather small, we can afford a loose bound for the operator $\|Q H_{BS} P_{1bpair}\|$, which is the other part of the integral. The argument is basically as follows. First, we notice that all operators in the product, $Q, H_{BS}$ and $P_{1bpair}$, commute with the total number of particles, which in our problem is exactly $N$. We can thus study the restrictions of these operators to this sector, which we denote as $P_N O P_N$ for each operator, where $P_N$ is the projector onto the space with $N$ particles. We then realize that $\|AB\|_2\leq \|A\|_2\|B\|_2$ and since the projectors have norm 1,
\begin{align}
  \|Q H_{BS} P_{1bpair}\|_2 &=\|Q P_N H_{BS} P_{N} P_{1bpair}\|_2 \\
  &\leq \|Q \|_2 \|P_N H_{BS} P_{N}\|_2 \|P_{1bpair} \|_2 \nonumber\\
  &=\|P_N H_{BS} P_{N}\|_2 \nonumber
\end{align}
\begin{figure}[t!]

  \centering\includegraphics[width=\linewidth]{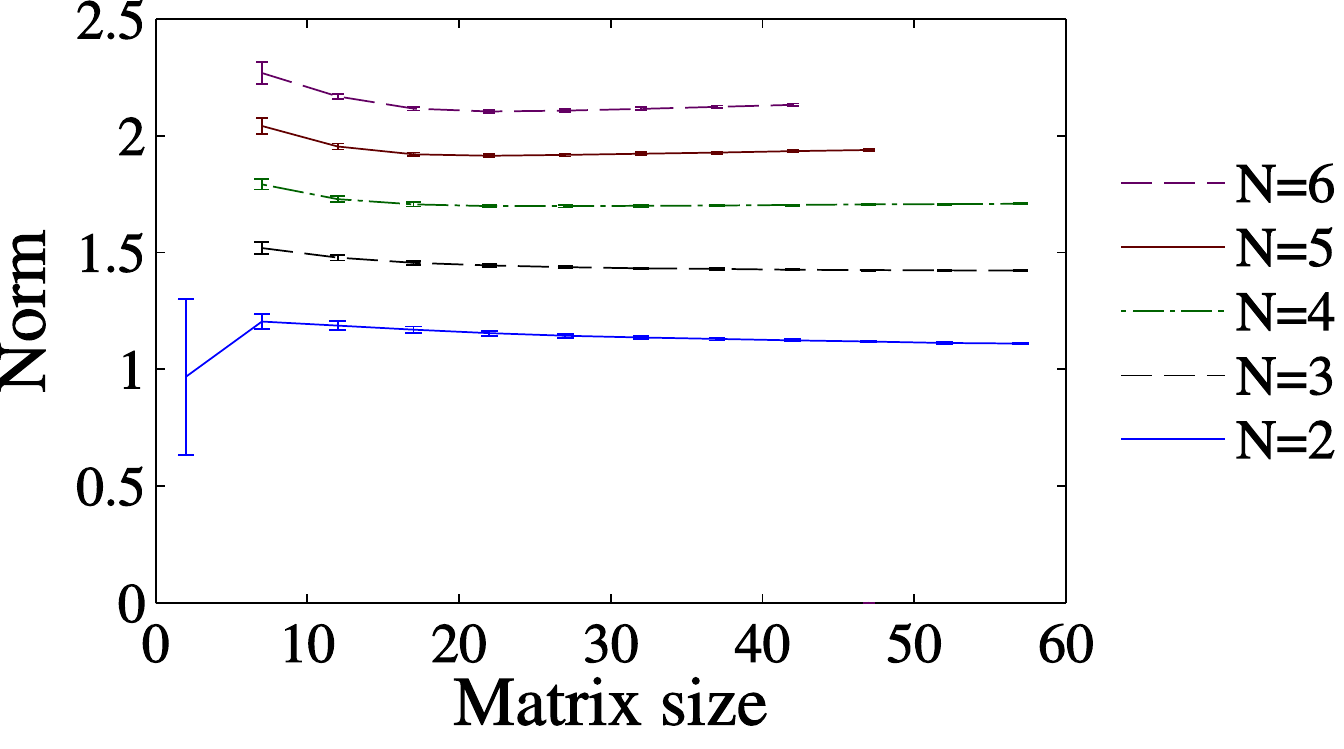}
  \caption{Numerical estimates of the norm $\|Q H_{BS} P_{1bpair}\|_2$ as a function of the number of bosonic modes, $M$, for different number of excitations, $N$.}
  \label{fig:norm}
\end{figure}
Notice now that $P_NH_{BS}P_N$ is just the Hamiltonian of $N$ free bosons, without hard-core restrictions of any kind. In other words, it is the restriction of
\begin{equation}
  H_{BS} = \sum_k (b_k^\dagger c_k + \mathrm{H.c.})
\end{equation}
to a situation where $\sum_k c_k^\dagger c_k + b_k^\dagger b_k = N$. We introduce superposition modes, $\alpha_{k\pm}=(c_k\pm b_k)/\sqrt{2}$, and diagonalize
\begin{equation}
  H_{BS} = \sum_k (\alpha_{k+}^\dagger \alpha_{k+} - \alpha_{k-}^\dagger \alpha_{k-}),
\end{equation}
where the constraint is the same $\sum_k \alpha_{k+}^\dagger \alpha_{k+} + \alpha_{k-}^\dagger \alpha_{k-} = N$. Since the largest eigenvalues (in modulus) are obtained by filling $N$ of these normal modes with the same frequency sign, we have
\begin{equation}
  \|Q H_{BS} P_{1bpair}\|_2 \leq \|P_N H_{BS} P_{N}\|_2 = N .
\end{equation}

Note that this proof does not make use of any properties of $H$ such as the fact that it is built from random matrices. As explained in the body of the letter, if we sample $Q H_{BS} P_{1bpair}$ randomly with the Haar measure and average the resulting norms, the bound seems closer to ${\mathcal O}(\sqrt{N})$.

We have strong evidence that this bound can be significantly improved using the properties of random matrices $R_{ij}$ and the structure of $Q H_{BS}\varepsilon$. In particular, we have numerical evidence that the average norm over the Haar measure is $\|Q H_{BS} P_{1bpair}\|_2 \propto \mathcal{O}(N^{1/2})$, which improves the requirement for efficient spin sampling $N\sim \mathcal{O}(M^{1/3})$. Fig.\ \ref{fig:norm} shows the average and standard deviation of the operator norm obtained by sampling random bosonic circuits with $N=2-6$ particles in $M=7-60$ modes, creating random unitaries according to the Haar measure and estimating the norm of the operator $Q H_{BS}P_{1bpair}$ with a sparse singular value solver. Note how, despite the moderate sample size (200 random matrices for each size) the standard deviation is extremely small, indicating the low probability of large errors and the efficiency of the sampling.

{ 
\section{Variation distance}
Throughout this manuscript, we have found bounds according to the 2-norm, in contrast to Aaronson and Arkhipov work, whose results are expressed in terms of the variation distance between probability distributions (this is, 1-norm). However, our proof above can be written in a similar was as the one given by Aaronson and Arkhipov, that is
\begin{equation}
|p_1-p_2|_1 := \sum_n |p_1(\mathbf{n})-p_2(\mathbf{n})|,
\label{vardis}
\end{equation}
which represents total difference between probabilities for all configurations $\mathbf{n}=(n_1,\ldots, n_M)$ of the occupations at the output ports.
In our model, the probability distribution associated to boson sampling would be
\begin{equation}
p_1(\mathbf{n}) = \left|\braket{\mathbf{n}|\phi(t)}\right|^2 =:|\phi(\mathbf{n})|^2,
\end{equation}
where if we focus on events with $n_i\in\{0,1\}$, we can replace $\phi$ with $Q\phi$. The corresponding probability for the spin model would be
\begin{equation}
p_2(\mathbf{n}) = \left|\braket{\mathbf{n}|\psi(t)}\right|^2=:|\psi(\mathbf{n})|^2,
\end{equation}

Using the above expressions for the probability distributions, we can write down the following identities for the total variation distance (\ref{vardis}).
\begin{eqnarray}
&&\sum_n |p_1(\mathbf{n})-p_2(\mathbf{n})|=\sum_n ||{\psi(\mathbf{n})}|^2-|{\phi(\mathbf{n})}|^2|\nonumber\\
&=& \sum_n |\psi^*(\mathbf{n})\psi(\mathbf{n})-\phi^*(\mathbf{n})\phi(\mathbf{n})|\\
&=&\frac{1}{2}\sum_n |[\left\langle{\psi(\mathbf{n})+\phi(\mathbf{n})}|{\delta(\mathbf{n})}\right\rangle+\left\langle\delta(\mathbf{n})|\psi(\mathbf{n})+\phi(\mathbf{n})\right\rangle]|\nonumber,
\end{eqnarray}
where $\ket{\delta(\mathbf{n})}=\ket{\psi(\mathbf{n})-\phi(\mathbf{n})}$. Hence, it follows that
\begin{eqnarray}
&\sum_n& ||{\psi(\mathbf{n})}|^2-|{\phi(\mathbf{n})}|^2|=\sum_n |\text{Re}(\left\langle{\psi(\mathbf{n})+\phi(\mathbf{n})}|{\delta(\mathbf{n})}\right\rangle)|\nonumber\\
&=&\sum_n |\text{Re}(2\phi^*(\mathbf{n})\delta(\mathbf{n})+\delta^*(\mathbf{n})\delta(\mathbf{n})|\nonumber\\
&\leq& 2\sum_n |\phi(\mathbf{n}) ||\delta(\mathbf{n}) |+\sum_n|\delta (\mathbf{n})|^2\\\
&\leq& 2 \left (\sum_n |\phi(\mathbf{n}) |^2 \right )^{1/2}\left (\sum_n |\delta(\mathbf{n}) |^2 \right )^{1/2}+\Vert \delta \Vert_2^2\nonumber\\
&=&2\Vert \phi \Vert_2\Vert \delta\Vert_2+\Vert \delta \Vert_2^2,\nonumber
\end{eqnarray}
Since the boson sampling wavefunction is normalized, $\Vert\phi\Vert_2=1$, and $\Vert \delta
\Vert_2\leq 1$ we finally get the next tight bound for the variation distance
\begin{equation}
|p_1-p_2|\leq 3\Vert\delta\Vert_2 = \mathcal{O}\left(\frac{N^2}{\sqrt{M}}\right).
\end{equation}}

\end{document}